\documentclass{aa501}
\usepackage{graphicx, epsfig, lscape}
\begin{document}
%
%\thesaurus{02     % A&A Section 2: Cosmology
%              (11.03.1;  % Galaxies: clusters: general,
%              11.03.4;  % Galaxies: clusters: individual: ...,
%              12.12.1)} % (Cosmology:) large-scale structure of Universe.

   \title{The velocity field of collapsing spherical structures}

   \subtitle{Limitations of the spherical infall model in mass estimation}

   \author{N. Hiotelis}
   \thanks{\emph{Present address:} Roikou 17-19,
               Neos Kosmos, Athens, 11743 Greece}

   \offprints{N. Hiotelis}

   \institute{Lyceum of Naxos, Chora of Naxos, Naxos 84300, Greece\\
              email: hiotelis@avra.ipta.demokritos.gr}

\date{Received 15 January 2001 / Accepted 28 May 2001 }

\abstract{ We assume that the amplitude of the caustics in
redshift space is a sum of two components: the first one can be
predicted by the spherical infall model with no random motion, and
the second is due to the random motion distribution. Smooth model
curves are used to estimate the maximum values of the first
component for the Coma cluster. Then, an approximation of the
radial component of the infall velocity --based on the above
curves-- is derived and a mass profile of the cluster is
calculated. This mass profile, that is an upper limit for the
spherical infall model, combined with estimations given by other
authors provides an approximation of a lower limit for the mass of
the system.
}

\maketitle
%
%________________________________________________________________

\section{Introduction}

The spherical infall model has been extensively discussed in the literature (Gunn \&
Gott \cite{gunn72}; Silk \cite{silk}; Gunn \cite{gunn77}; Peebles (\cite{peeb76},
\cite{peeb80}); Schechter \cite{schechter}; Schectman \cite{shectman}; Ostriker et
al. \cite{ostriker}). This model assumes that galaxy clusters started as small
density perturbations in the early universe. These perturbations eventually deviate
from the general expansion and after reaching a maximum radius they start
collapsing. The most probable scenario is that galaxies formed during the expansion
phase of their cluster. Thus, the cluster consists of largely individual galaxies
during its collapse.

The observed velocities of these galaxies (along the
line--of--sight) give important information about the dynamical
state of the cluster (Kaiser \cite{kaiser}). Reg\"os \& Geller
(\cite{regos}) showed that plotting the observed velocities of
galaxies as function of their angular distances from the centre of
the cluster one obtains a velocity distribution that is bounded by
sharp, characteristic trumpet--shaped curves. These curves, called
caustics, form an envelope containing the galaxies of the cluster.
The caustics are used to estimate the density parameter
$\Omega_{0}$ of the universe (Reg\"os \& Geller \cite{regos}) and
the mass profile for the galaxy clusters (Geller et al.
\cite{gell99}; Reisenegger et al. \cite{reinse}). In Sect.~2, we
show that the form of the caustics fully defines the profile of
the radial infall velocity in the case of a pure spherical
collapse. In Sect.~3, we describe the spherical model. This model
is applied in Sect.~4 to data from the Coma cluster. In Sect.~5,
the conclusions are given.
%
%                                                One column figure
%----------------------------------------------------------- S_vib
\begin{figure}[b] \hskip 0.30 true cm
\includegraphics[angle=-90,width=8.0cm]{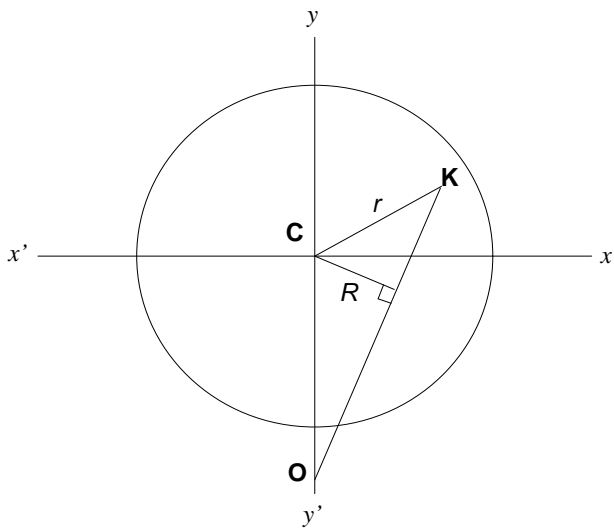}
\caption{Geometry of the problem. O: observer, K: galaxy, C: cluster's centre}
\label{fig1}
\end{figure}
%-----------------------------------------------------------
%

\section{Observed velocities}

For the sake of simplicity, let us consider the $xy$ plane of a
spherically symmetric cluster, undergoing a radial collapse
(Fig.\ref{fig1}), where $\vec{r}$ is the position vector of a
galaxy at K, relative to the centre C of the cluster. The velocity
$\vec{v}$ of the galaxy at K relative to the observer is given by
the relation \setcounter{equation}{0}
\begin{equation}
\vec{v} = \frac{\mathrm{d}\vec{r}}{\mathrm{d}t}+ \vec{v}_{\mathrm{C}}, \label{eqb1} \\
   \end{equation}
\noindent where $\vec{v}_{\mathrm{C}}$ is the velocity of the centre C relative to
O. Defining $v_{r}(r)= \mathrm{d}r / \mathrm{d}t$, where $r$ is the magnitude of
$\vec{r}$ and multiplying both sides of eqn.(\ref{eqb1}) by the unit vector
$\hat{\vec{r}}_\mathrm{K}$ along \overrightarrow{OK} direction, we have\

\begin{equation}
 v_{\mathrm{obs}}(r,R) = \vec{v}_\mathrm{C} \cdot \hat{\vec{r}}_\mathrm{K} \pm
\sqrt{1-\left({\frac{R}{r}}\right)^{2}}v_{r}(r),\label{eqb2} \\
\end{equation}

\noindent where $v_{\mathrm{obs}}=\vec{v} \cdot
\hat{\vec{r}}_\mathrm{K}$ is the observed velocity along the
line--of--sight.
  From eqn.(\ref{eqb2}), for $r=R$ and
for $r=r_\mathrm{M}$, the observed velocity takes the same value
$v_{\mathrm{obs}}(r,R) = \vec{v}_\mathrm{C} \cdot
\hat{\vec{r}}_\mathrm{K}$. Thus, $v_{\mathrm{obs}}$ takes extreme
values for $r$ between $R$ and $r_\mathrm{M}$. We denote by
$r_\mathrm{M}$ the radius of maximum expansion (turnaround radius)
where $v_{r} (r_\mathrm{M})=0$.

For constant $R$, the quantity $\vec{v}_\mathrm{C} \cdot
\hat{\vec{r}}_\mathrm{K}$ does not depend on $r$ and thus the
condition $\partial v_{\mathrm{obs}} /
\partial r =0$ for the extreme is written

\begin{equation}
\left[1-\left({\frac{R}{r}}\right)^{2}\right]\frac{\mathrm{d}
v_{r}(r)}{\mathrm{d}r}+ \frac
{R^{2}}{r^{3}}v_{r}(r)=0. \label{eqb3} \\
\end{equation}

 \noindent The solution $r_{*}$ for $r$ of eqn.(\ref{eqb3})
  is a function of $R$ ($r_{*}=r_{*}(R)$). This $r_{*}$ is the
radial --relative to the centre of the cluster-- distance of the
galaxy which shows the observer the maximum or minimum velocity at
a given $R$. Thus, the extreme values of $v_{\mathrm{obs}}$ at $R$
are

\begin{equation}
v_{\mathrm{obs,ext}}(R)=\vec{v}_{\mathrm{C}} \cdot
{\hat{\vec{r}}}_\mathrm{K} \pm
\sqrt{1-\left[{\frac{R}{r_*(R)}}\right]^{2}}v_{r}(r_*(R)). \label{eqb4} \\
\end{equation}

\noindent The curves described by eqn.(\ref{eqb4}) are known as caustics (Reg\"os \&
Geller (\cite{regos})).

Following Diaferio et al. (\cite{diaf97}), we define the amplitude
$A$ of the caustics by the relation

\begin{equation}
A(R)= \frac {1}{2} [v_{\mathrm{obs,max}}(R)-
v_{\mathrm{obs,min}}(R)],
\end{equation}

\noindent which is given by

\begin{equation}
A(R)= -
\sqrt{1-\left({\frac{R}{r_*}}\right)^{2}}v_{r}(r_*).\label{eqb6} \
\end{equation}

\noindent Note that by definition $v_r$ is negative in the infall region of the
cluster. The amplitude contains only information about the dynamical situation of
the cluster, since effects due to the motion of the observer relative to the cluster
are excluded. Such effects are described by the term $\vec{v}_\mathrm{C} \cdot
\hat{\vec{r}}_\mathrm{K}$ in eqn.~(\ref{eqb4}) and have been studied by Praton \&
Schneider (\cite{praton}).
 Differentiating
eqn.(\ref{eqb6}) with respect to $R$, we have

\begin{eqnarray}
 \frac {\mathrm{d}A}{\mathrm{d}R} =\frac {R v_r(r_*)} {r_*^{2}
\sqrt{1-\left({\frac{R}{r_*}}\right)^{2}}}\nonumber\\
- \left[\frac{R^{2}v_r(r_*)} {r_*^{3}
\sqrt{1-\left({\frac{R}{r_*}}\right)^{2}}} +
\sqrt{1-\left({\frac{R}{r_*}}\right)^{2}} \frac {\mathrm{d}v_r
(r_*)} {\mathrm{d}r_*} \right ]\frac {\mathrm{d}r_*}
{\mathrm{d}R}.\label{eqb7}
\end{eqnarray}

\noindent Because of eqn. (\ref{eqb3}), the term inside brackets
equals zero and this leads to

\begin{equation}
\frac {v^{2}_r (r_*)} {r_*^{2}}= - \frac {A(R)} {R} \frac
{\mathrm{d}A(R)} {\mathrm{d}R}.\label{eqb8} \
\end{equation}

\noindent Using eqns. (\ref{eqb6}) and ~(\ref{eqb8}) we have

\begin{equation}
r_*=R \sqrt {1-\left ( \frac {\mathrm{d}\ln A(R)} {\mathrm{d}\ln
R}\right )^{-1}}.\label{eqb9} \
\end{equation}

\noindent Solving eqn.(\ref{eqb9}) for $R$ and substituting in eqn.(\ref{eqb8}) the
radial velocity is fully defined. This is an interesting result, since it shows that
the exact data can give the exact profile of the infall velocity, in the case of a
completely radial collapse. It is interesting to note that the above relations lead
to eqn.({\ref{eqb10}), relating the logarithmic derivatives of the amplitude to the
infall velocity;\

\begin{equation}
\frac {\mathrm{d}\ln [-v_{r}(r_{*})]}{\mathrm{d}\ln r_{*}}= \frac
{\mathrm{d}\ln A(R)}{\mathrm{d}\ln R}\label{eqb10} \
\end{equation}

\noindent The above equations refer to the ideal case of a pure spherical collapse
and completely accurate observations. In real systems, the velocity field is a
superposition of a radial systematic component and a component of a random nature.
The first one can be assumed as spherically symmetric while the second accounts for
the effects of small-scale substructure and observational errors. The effects of
small-scale substructure are clearly shown in the results of N-body simulations
performed by van Haarlem and van de Weygaert (\cite{van}). They conclude that the
velocity profile, as predicted by the spherical infall model, compares badly with
the actual velocity field resulting from their simulations. In real systems, such as
the Coma cluster, the smearing of the form of the caustics is also clear (van
Haarlem et.al \cite{vh}). However, if the systematic radial component of the
velocity field is known, a mass profile of the infall regions of the clusters can be
determined, applying the spherical infall model. This profile should give a lower
limit for the mass of the system since the effects of small-scale substructure
reported above increase the amplitude of the caustics. Thus, we apply the previous
eqns. assuming that $v_r$ describes the systematic radial component of the velocity.
In Sect.4 we use model functions for the amplitude $A$, calculating  the profile of
$v_r$ by eqns. (\ref{eqb8}) and (\ref{eqb9}), we apply the spherical infall model
and we derive a mass profile for the Coma cluster. These model functions for $A$ are
smooth and decreasing, as it required by eqns. (\ref{eqb8}) and (\ref{eqb9}), and
their approximation is based on the respective curves estimated by Geller et al.
(\cite{gell99}).\
\section{The spherical infall model~--~Mass estimation}
The eqn. of motion of a spherical shell of initial radius $a$ is given by Newton's
law as
\begin{equation}
\ddot{r}=\frac{\mathrm{d}^2r}{\mathrm{d}t^2}=-\frac{GM(r)}{r^2}\label{eqb11}
\end{equation}
\noindent where $M(r)$ is the mass inside radius $r$ and $G$ is the gravitational
constant. If the mass inside the shell is conserved (no shells crossing), then
$M(r)$ is constant and equal to $M(a)$. We assume that this condition holds in what
follows. Multiplying eqn.(\ref{eqb11}) by $\dot{r}$ and integrating, we have
\begin{equation}
\left(\frac{\mathrm{d}r}{\mathrm{d}t}\right)^2=v^2(t)=
2E+\frac{2GM(a)}{r}
\end{equation}
\noindent where $E={1}/{2}~{v_i}^2-(GM(a))/a$ is a constant with
dimensions of specific energy and $v_i$ is the initial velocity.
The solution of the above eqn., in the case of negative $E$, is
given in parametric form by the expressions\

\begin{equation}
  r=\frac{GM(a)}{(-2E)}(1-cos\eta)\label{eqs1}\
\end{equation}

\begin{equation}
  t=\frac{GM(a)}{(-2E)^{\frac{3}{2}}}(\eta-sin\eta)+t_a.\label{eqs2}\
\end{equation}
\noindent In order to synchronize all shells, we add the constant $t_a$ given by
$t_a=t_i-\frac{GM(a)}{(-2E)^{\frac{3}{2}}}(\eta_i-\sin{\eta_i})$, where $t_i$ is the
age of the universe at the initial conditions and $\eta_i$ is the initial phase of
the shell, given by eqn.(\ref{eqs1}) for $r=a$. A shell reaches the radius of
maximum expansion for $\eta=\pi$ and completes its collapse for $\eta=2\pi$. We
denote by $\mu(a)$ the fractional excess of mass of a sphere of radius $a$, relative
to the mass $M_b(a)$ of a sphere of the same radius that has a constant density
(equal to the mean density of the universe at the initial conditions). Using the
relation between the mean density of the universe $\rho_b$, the Hubble constant $H$
and the density parameter $\Omega$, that is
\begin{equation}
\rho_b(t)=\frac{3H^2(t)\Omega(t)}{8\pi G},
\end{equation}
\noindent the mass inside radius $a$ is given by
\begin{equation}
M(a)=\frac{H^2_i\Omega_ia^3}{2G}(1+\mu(a)),\label{eqb16}\
\end{equation}

\noindent where index $i$ stands for the values at the initial
conditions. Assuming that the initial velocity of the shell is
that of Hubble's flow ($v_{i}=aH_{i}$), the energy $E$ is written\
\begin{equation}
E=\frac{1}{2}H^2_ia^2\Omega_i\left[{\Omega_i}^{-1}-1-\mu(a)\right]
\label{eqb17}\
\end{equation}
\noindent Substituting eqn.(\ref{eqb17}) in eqns. (\ref{eqs1}),~(\ref{eqs2}) and
calculating the time derivative of the radius yields the relations
\begin{equation}
r=\frac{a}{2}\frac{{\mu_e(a)}
+{\Omega_i}^{-1}}{\mu_e(a)}(1-\cos{\eta})\label{eqb18}\\
\end{equation}
\begin{equation}
t=\frac{\mu_e(a)+{\Omega_i}^{-1}}{2H_i{\Omega_i}^{\frac{1}{2}}
{\mu_e}^{\frac{3}{2}}(a)}(\eta-\sin{\eta})+t_a\label{eqb19}\\
\end{equation}
\begin{equation}
v_r=aH_i{\Omega_i}^{\frac{1}{2}}{\mu_e}^{\frac{1}{2}}
(a)\frac{\sin{\eta}}{1-\cos{\eta}}\label{eqb20}\\
\end{equation}
\noindent where $\mu_e(a)=1+\mu(a)-\Omega_i^{-1}$.\

\noindent Combining eqns. (\ref{eqb18}), (\ref{eqb19}) and (\ref{eqb20}) results in
\begin{equation}
(t-t_a)\frac{v_r}{r}=\frac{\sin{\eta}(\eta-\sin{\eta})}
{(1-\cos{\eta})^2},\label{eqb21}\\
\end{equation}
\noindent where $t$ is the present age of the shell. The time $t$
is approximated by $t={H_0}^{-1}f(\Omega_{0},z)$, where $z$ is the
redshift of the cluster,  $H_0$ is the present value of the Hubble
constant and $\Omega_0$ is the value of the density parameter. The
function $f$ for the Friedmann-Lemaitre cosmological model is
given in standard books of cosmology (see Zel'dovich \& Novikov
{\cite{zeldovich}) and its form depends on the value of $\Omega_0$
relative to the critical value $\Omega_{0}=1$. Reasonable initial
conditions are those at the epoch of decoupling, when matter
starts playing a significant role. The value of $t_a$ is very
small compared to $t$, so it can be omitted without significant
error in the calculation of the phase $\eta$. However,
 eqn. (\ref{eqb21}) can also be solved in its full form based on
the following procedure. Solving eqn.(\ref{eqb18}) for $\mu_e$
gives
\begin{equation}
\mu_e(a)=\frac{a(1-\cos{\eta}){\Omega_i}^{-1}}{2r-a(1-\cos{\eta})} .
\end{equation}
\noindent The substitution of the above expression in eqn.(\ref{eqb20}) leads to the
following cubic for $a$
\begin{equation}
a^3+(1-\cos{\eta})^2Ca-2r(1-\cos\eta)C=0
\end{equation}
\noindent where $C=v^2_r/(\sin^2{\eta}H^2_i)$. This has one real
and positive root, given by the relation
\begin{equation}
a=\left(\frac{rv^2_r}{H^2_i(1+\cos\eta)}\right)^{\frac{1}{3}}
\left[(1+\sqrt{F})^{\frac{1}{3}}+(1-\sqrt{F})^{\frac{1}{3}}\right]
\end{equation}
\noindent where
\begin{equation}
F=1+\frac{1}{r^2}\frac{(1-cos\eta)^3}{27} \frac{v^2_r}{H^2_i(1+\cos\eta)}.
\end{equation}
\noindent Thus, $\mu_e$ can be expressed as a function of
$r,v_r,H_i,\Omega_i$ and $\eta$. Then, $t_a$ can also be expressed
in terms of the same variables, since it is given by the relation
\begin{equation}
t_a=t_i- \frac{1}{2H_i\Omega^{\frac{1}{2}}_i}\frac{\mu_e(a)+
\Omega^{-1}_i}{\mu_{e}^{\frac{3}{2}}(a)}
(\varphi-{\sin}^{-1}\varphi)\label{eqb26}
\end{equation}
\noindent where $\varphi=\frac{2\mu^{\frac{1}{2}}_e(a)}
{\Omega^{\frac{1}{2}}_i(\mu_e(a)+\Omega^{-1}_i)}$\ and $t_i$ is
the age of the universe at the initial conditions (decoupling),
given in standard books of cosmology (see Zel'dovich \& Novikov
{\cite{zeldovich}).\

The present method can be summarized as follows: A known profile of the amplitude of
the caustics, using eqns. (\ref{eqb8}) and (\ref{eqb9}) leads to the profile of
radial velocity. Then, the solution of eqn.(\ref{eqb21}) defines the phase $\eta$ of
any shell. Assuming no shell crossing during the evolution of the cluster, the
combination of eqns. (\ref{eqb16}),~(\ref{eqb18}) and (\ref{eqb20}) gives the mass
inside a spherical region with current radius $r$ as
\begin{equation}
M(r)=\frac{r{v^2}_r}{G}\frac{1-\cos{\eta}}{sin^2\eta}.
\end{equation}
\section{Applications}
The procedure described above is applied to estimate a lower limit
for the mass profile of the Coma cluster. For this purpose, we
employ data provided by Geller et al. (\cite{gell99}). The largest
sample presented by these authors contains 691 galaxies. The
cluster's velocity $cz_\mathrm{C}$ ($c$ is the speed of light and
$z_\mathrm{C}$ is the redshift of the cluster centre) is
$7\,090\mathrm{Km\,sec^{-1}}$ and for a Hubble constant
$H_0=100\mathrm{h\,Km\,sec^{-1}\,Mpc^{-1}}$, its distance is
$70.9\mathrm{h^{-1}\,Mpc}$.

\begin{figure}[b]
\includegraphics[width=8.6cm]{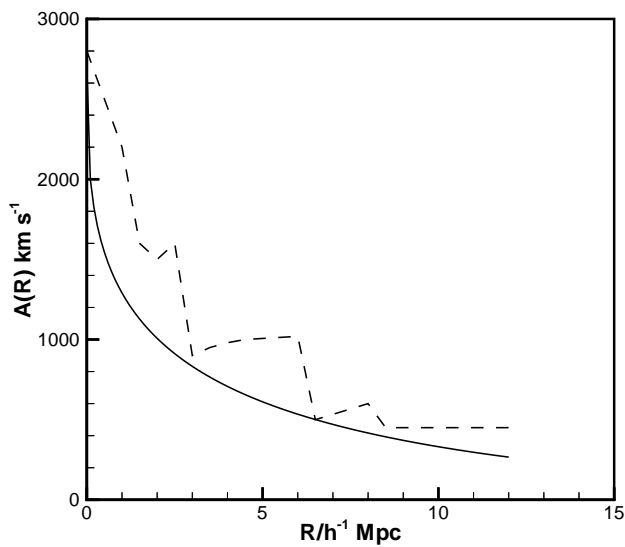}
\caption{ Dashed line: profile of the amplitude of the caustics (Geller et al.
(\cite{gell99})). Solid curve: approximation of this line using analytical smooth
functions (see eqn.(\ref{ab1}) and text for more details.)} \label{fig2}
\end{figure}

As noted in Sect. 2, our approach is based on the profile of the amplitude $A$ of
the caustics. Departures from spherical symmetry, random motions due to the
development of substructure, the finite number of galaxies in the cluster and
observational errors are some of the reasons making the determination of caustics
non-trivial. A general approach, given by Diaferio (\cite{diaf99}), is based on the
argument that caustics are high density regions (see Reg\"{o}s \& Geller
\cite{regos}) and is as follows. First, a smooth estimate for the density $f$ of
observed galaxies on the $(R,v_{obs})$ plane is calculated and then a cut is applied
at some density contour which is taken to correspond to the caustics. A similar
procedure is followed by Reisenegger et al. (\cite{reinse}) in their application to
the Sharpley Supercluster. Problems associated with the above procedure, such as the
choice of smoothing lengths and the density cutoff, are discussed in the above
papers. The dashed line --shown in Fig.~\ref{fig2}-- is Geller et al.'s
(\cite{gell99}) estimation of the amplitude $A_{est}(R)$ of the caustics derived by
the above--discussed procedure. We assume that this can be written in the form
\begin{displaymath}
A_{est}(R)=A(R)+dA(R)
\end{displaymath}
\noindent where $A(R)$ stands for the component due to the radial
infall velocity and the positive $dA(R)$ describes the

%-----------------------------------------------------------
%
%                                                One column figure
%----------------------------------------------------------- S_vib
\begin{figure}[t] \hskip 0.30 true cm
\includegraphics[width=8.6cm]{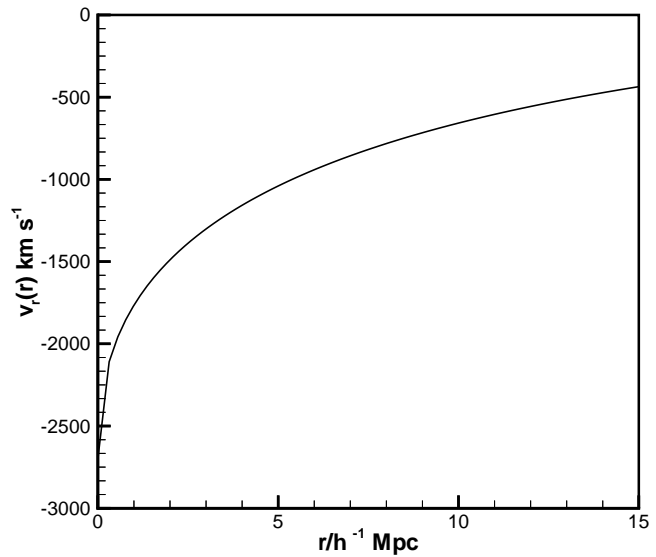}
\caption{Radial velocity profile corresponding to the
      smooth curve plotted in Fig.\ref{fig2}}\label{fig3}
\end{figure}

\noindent 'inflation' of the caustics due to the development of
random motions. Unfortunately, $dA(R)$ cannot be modelled properly
and an infinite number of curves lower than $A_{est}$ could be
provided as  models for $A$. For this reason we use the following
procedure. First, we use models for  $A(R)$ of the form
\begin{equation}
A(R)=v_{0}\left[1-\left(\frac{R}{r_{\mathrm M}}\right)^{\frac{1}{k}}\right]^k,~
v_{0}\geq 0,~R\leq r_{\mathrm M}. \label{ab1}
\end{equation}
Then, we estimate the values of the parameters $v_{0},r_{M}$ and
$k$ in order that these curves satisfy the condition $dA(R)>0$ for
all values of $R$ and minimize the sum $\sum [A_
{est}(R)-A(R)]^2$. In this way, we derive a profile that could be
considered as an upper bound for the $A$ component.
 \noindent Such a curve is plotted in
Fig.\ref{fig2}. The parameters used in eqn.(\ref{ab1}) are:
 $v_0=2\,700\mathrm{Km\,sec^{-1}}, r_{\mathrm
 M}=60h^{-1}\mathrm{Mpc},k=2.8$. We note that $r_{\mathrm M}$ is not the
physical radius of the system, but a fitting parameter. The form of eqn.(\ref{ab1})
has mathematical advantages, since it allows an analytical evaluation of the profile
of the infall velocity, using eqns. (\ref{eqb8}) and (\ref{eqb9}) reported in
Sect.~2. This is given by the relation
\begin{equation}
v_{r}(r)=-v_{0}\left[1-\left(\frac{r}{r_{\mathrm M}}\right)^
{\frac{2}{2k-1}}\right]^{k-{\frac{1}{2}}},~r \leq {r_{\mathrm
M}},~k>{\frac{1}{2}}\
\end{equation}

\noindent with
$r=\left(\frac{r_\mathrm{M}}{R}\right)^{\frac{1}{2k}}R$.

In Fig.\ref{fig3}, the resulting profile of the radial infall
velocity  is plotted.
 Assuming that the above profile is a good approximation
  of the radial systematic component of the infall velocity, we
 apply the spherical infall model to calculate mass profiles, for various values of the density
 parameter $\Omega_0$ of the universe,  that are plotted
 in Fig.\ref{fig4}. Starting from the bottom of the figure, the different curves correspond to
$\Omega_0=0.2,0.4,0.6,0.8,1.0$ respectively. Larger values of
$\Omega_0$ lead to steeper mass profiles at the outer regions. The
variation of $\Omega_0$ alters the profile at the outer regions of
the cluster and (as expected) larger values of the density
parameter of the universe give rise to larger amount of mass
inside a given radius. Notice that the mass scales as
${\Omega_0}^{0.2}$, roughly as expected (see Reg\"{o}s \& Geller
\cite{regos}). Thus, the matter inside $10h^{-1}\mathrm{Mpc}$
varies for various values of the density parameter of the universe
in the range $0.994$ to $1.12\times 10^{15}h^{-1}M_{\sun}$. The
dashed curve, plotted also in this Fig., is the cumulative mass of
an halo with a Navarro et al. (\cite{navarro}) profile. This is
given by the relation
$M(r)=4\pi{\rho}_0{r^3}_s\{\ln[1+(r/rs)]-(r/rs)/[1+(r/rs)]\}$,
where $\rho_0$ and $r_s$ are fitting parameters. The values of the
parameters used are: $\rho_0=3.8\times10^{15}h^{2}
M_{\sun}\mathrm{Mpc}^{-3}$, $r_s=0.2h^{-1}\mathrm{Mpc}$. It is
clearly shown in Fig.\ref{fig4} that there is a remarkable
agreement of the form of the cumulative mass profiles estimated by
the present approach (solid curves) with the Navarro et al.
(\cite{navarro}) mass profile (dashed curve). Note that the dashed
curve describes the mass profile in the virialized region of the
cluster.

\begin{figure}
\includegraphics[width=8.6cm]{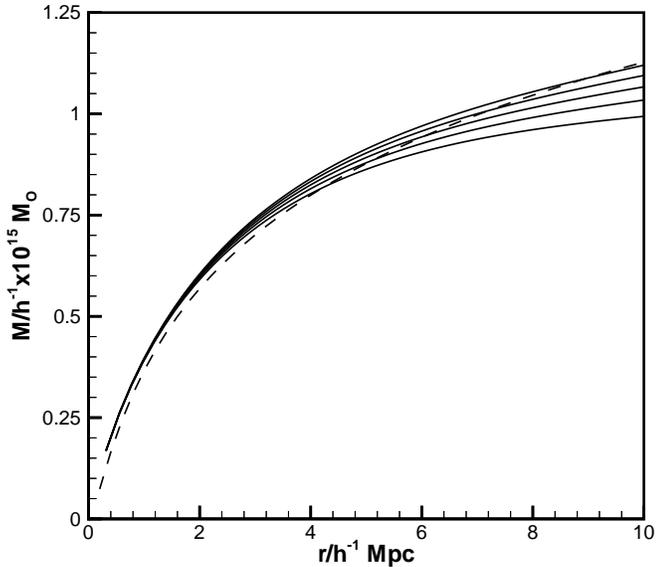}
\caption[]{Mass profiles corresponding
     to the radial velocity profile (Fig.\ref{fig3})
       for varying $\Omega_0$. From the bottom of the figure, the curves
      correspond to $\Omega_0=0.2,0.4,0.6,0.8,1.0$ respectively.
      Dashed curve: Navarro et al. (\cite{navarro})
       mass profile.}\label{fig4}
\end{figure}

\section{Conclusions}
\begin{figure}
\includegraphics[width=8.6cm]{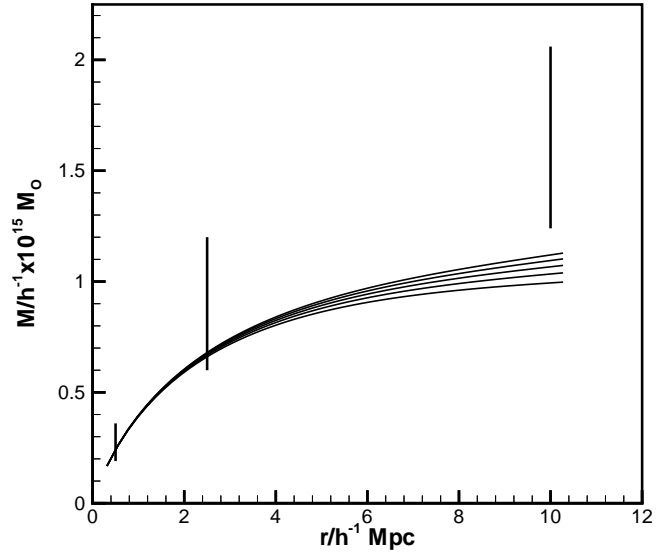}
\caption[]{Mass profiles for various values of
    $\Omega_0$ as in Fig.\ref{fig4}. The bar at $0.5$ shows the
    limits given by Hughes (\cite{hughes}) while the bar at $2.5$ shows the
    limits given by Briel et al. (\cite{briel}). The outer bar at
    $10$ shows the estimation given by Geller et al. (\cite{gell99})
      }\label{fig5}
\end{figure}
Several methods have been proposed in the literature regarding the
estimation of the mass of galaxy clusters.

A class of such methods, used to estimate the density profile of
these systems, is based on the use of X-ray data (The \& White
\cite{the}; Hughes \cite{hughes}; Watt et al. \cite{watt}). It is
assumed that the gas is described by the equation of hydrostatic
equilibrium. The latter is solved using the density and
temperature profiles of the gas, that are measured using X-ray
observations. This method is accurate in the region where the
hydrostatic equilibrium holds, that is the central region of
galaxy clusters. For example, Hughes (\cite{hughes}) determined
the minimum and maximum mass profiles consistent with all the data
for Coma --if the mass-to-light ratio of the data was not
constrained to be a constant-- obtaining allowed mass ranges of
$M_{0.5}=1.9-3.6\times10^{14}h^{-1}M{\sun}$ and
$M_{2.5}=0.5-1.5\times10^{15}h^{-1}M{\sun}$ where $M_{0.5}$ and
$M_{2.5}$ are the masses inside $0.5h^{-1}\mathrm{Mpc}$ and
$2.5h^{-1}\mathrm{Mpc}$ respectively. Briel et. al (\cite{briel})
used their ROSAT survey image to measure the X-ray surface
brightness out to $100'$ from the cluster center. They found that
the binding mass is more centrally concentrated than the X-ray
gas, and obtain $M_{2.5}=0.6-1.2\times10^{15}h^{-1}M{\sun}$.

Another class of methods employs the projected velocities
  of the galaxies-members of the
cluster. The methods based on the form of the caustics in redshift
space belong -among others- in this class. In a series of papers
(Diaferio et al. \cite{diaf97}; Diaferio \cite{diaf99}; Geller et
al. \cite{gell99}), it has been proposed to estimate the mass
profile of a galaxy cluster using the equation
\begin{equation}
GM(r)\approx\frac{1}{2}\int_0^r{A^2_{est}(x)\mathrm{d}x}.
\end{equation}
Although there is no rigorous proof for this relation, its testing
against cosmological N-body simulations shows that it approximates
well the mass profiles. The derivation of the above relation is
based on the argument that the amplitude of the velocity field of
galaxy clusters depends mainly on local dynamics, since the random
motion is significant. The mass profile is, in any case, roughly
proportional to the square of the amplitude of the caustics and
since the amplitude used by Geller at al. (\cite{gell99}) is
larger because it includes the effect of random motion, one
expects their estimation to result in a larger mass for the
system. They estimate a mass inside a radius of
$10h^{-1}\mathrm{Mpc}$ in the range $(1.65\pm0.41)\times
10^{15}h^{-1}M_{\sun}$.

In Fig.\ref{fig5}, we compare the mass estimations provided by our
method for the Coma cluster to the mass estimations given by
Hughes (\cite{hughes}) , Briel (\cite{briel}) and Geller et al.
(\cite{gell99}). It is clear that the mass profile resulting from
our application shows the minimum estimated values of mass at
different radii and provides a reasonable estimation of the lower
limit of the  mass profile of the system.

 More detailed
observations could lead to a more accurate definition of the
caustics and improve the estimation of mass profiles .

\begin{acknowledgements}
      Thanks to the EMPIRIKION Foundation for its support.
\end{acknowledgements}


\begin{thebibliography}{}

\bibitem[1992]{briel} Briel, U.G., Henry, J.P. \& B\"{o}hringer, H.
        1992, A\&A,259, L31

\bibitem[1997]{diaf97} Diaferio, A. \& Geller, M.J. 1997,
      ApJ 481, 633

\bibitem[1999]{diaf99} Diaferio, A. 1999,
      MNRAS 309, 610

\bibitem[1999]{gell99} Geller, M. J., Diaferio, A. \& Kurtz, M. J. 1999,
      ApJ 517, L23

\bibitem[1972]{gunn72} Gunn, J. E. \& Gott, J. R. 1972,
      ApJ 176, 1

\bibitem[1977]{gunn77} Gunn, J. E. 1977,
      ApJ 218, 592

\bibitem[1989]{hughes} Hughes, J.P. 1989,
ApJ 337,21

\bibitem[1987]{kaiser} Kaiser, N. 1987,
      MNRAS 227, 1

\bibitem[1995]{navarro} Navarro, J. F., Frenk, C. S. \& White, S. D. M.
 1995, MNRAS 275, 720

\bibitem[1988]{ostriker} Ostriker, E. C., Huchra, J. P., Geller, M. J., \& Kurtz, M.J. 1988,
      AJ 96, 1775

\bibitem[1976]{peeb76} Peebles, P. J. E. 1976,
      ApJ 205, 318

\bibitem[1980]{peeb80} Peebles, P. J. E. 1980, The Large Scale
Structure of the Universe (Princeton University Press, Princeton),
64

\bibitem[1994]{praton} Praton, E. A. \& Schneider, S. E. 1994,
      ApJ 422, 46

\bibitem[1989]{regos} Reg\"os, E. \& Geller, M. J 1989,
      AJ 98, 755

\bibitem[2000]{reinse} Reinsenegger, A., Quintana, H., Carrasco, E. R.,
et al. 2000, AJ 120, 523

\bibitem[1980]{schechter} Schechter, P. L. 1980,
      AJ 85, 801

\bibitem[1982]{shectman} Shectman, S. A. 1982,
      ApJ 262, 9

\bibitem[1974]{silk} Silk, J. 1974,
      ApJ 193, 525

\bibitem[1988]{the} The, L. S. \& White, S. D. M. 1988,
      AJ 95, 15
\bibitem[1993]{van}van Haarlem, M. P. \& van de Weygaert, R. 1993, ApJ, 418, 544

\bibitem[1993]{vh} van Haarlem, M. P., Cay\'{o}n, L., de la Cruz,
C.G., et al. 1993 MNRAS 264, 71

\bibitem[1992]{watt} Watt, M. P., Ponman, T.J., Bertram, D., et al.
1992, MNRAS 258, 738

\bibitem[1983]{zeldovich} Zel'dovich, Y. B. \& Novikov, I. D. 1983,
 The Structure and Evolution of the Universe
 (The University of Chicago Press, Chicago and London),
 54

\end{thebibliography}
\end{document}